\documentstyle[11pt,epsfig]{article}

\begin{document}

\begin{center}
{\Large 
\bf On the confidence interval for the parameter of Poisson distribution \\}
\end{center}

\bigskip

\begin{center}
{\large S.I.~Bityukov$^{a,}$\footnote{\small Corresponding author\\
{\it Email addresses:} bityukov@mx.ihep.su, Serguei.Bitioukov@cern.ch} 
N.V.~Krasnikov$^b$, V.A.~Taperechkina$^c$}\\  

$^a${\it {\footnotesize Division of Experimental Physics,
Institute for High Energy Physics, Protvino, Moscow Region, Russia}}\\
$^b${\it {\footnotesize Division of Quantum Field Theory,
Institute for Nuclear Research RAS, Moscow, Russia}}\\
$^c${\it {\footnotesize Mathematics and Computer Science Department,
Moscow State Academy of Instrument Engineering and Computer Science,
Serpukhov, Moscow Region, Russia}}\\

\end{center}


\begin{flushleft}
{\large \bf Abstract}\\

\bigskip

The possibility of construction of continuous analogue of 
Poisson distribution with the search of bounds of confidence 
intervals for parameter of Poisson distribution is discussed. 
Also, in the article is shown that the true value of a parameter of Poisson 
distribution for the observed value $\hat x$ has Gamma distribution 
with the scale parameter, which is equal to one, and the shape parameter,
which is equal to $\hat x$. The results of numerical construction 
of confidence intervals are presented.

\bigskip

\end{flushleft}


\section{Introduction}

In article~\cite{1} the unified approach to the construction
of confidence intervals and confidence limits
for a signal with  a background presence, 
in particular for Poisson distributions, is proposed. The method is widely
used for the presentation of physical results~\cite{2}
though a number of investigators criticize this approach~\cite{3}
(in particular, this approach avoids a violation of the coverage principle).
Series of Workshops on Confidence Limits has been held in CERN and
Fermilab. At these meetings demands for properties of constructed
confidence intervals and confidence limits have been formulated~\cite{4}.
On the other hand, the results of experiments often give noninteger
values of a number of observed events 
(for example, after background subtraction~\cite{5})
when a Poisson distribution
take place. That is why there is a necessity to search
a continuous analogue of Poisson distribution. The present work offers
some generalization  of Poisson distribution for continuous case.
The generalization given here allows to construct 
confidence intervals and confidence limits
for Poisson distribution parameter both for integer and real values
of a number of observed events, using conventional methods.
More than, the supposition about continuous of some function $f(x,\lambda)$
described below allows to use a Gamma distribution for construction of
confidence intervals and confidence limits of Poisson distribution parameter.
In present article we consider 
only the construction of confidence intervals.

In the Section 2 the generalization of Poisson distribution for the
continuous case is introduced. An example of confidence intervals
construction for the parameter of analogue of Poisson distribution is given 
in the Section 3. In the Section 4 the results of construction of
confidence intervals having the minimal length for the parameter of
Poisson distribution using a Gamma distribution are discussed.
The main results of the article are formulated in the Conclusion.

\section{The Generalization of Discrete Poisson Distribution for
the Continuous Case}

Let us have a random value $\xi$, taking values from the set of
numbers $x \in X$. Let us consider two-dimensional function

\begin{equation}
f(x,\lambda) =\displaystyle \frac{\lambda^x}{x!} e^{-\lambda}, 
\end{equation}

\noindent
where $x \ge 0$ É $\lambda > 0$.
Assume, that set $X$ includes only integer numbers, then discrete
function $f(x,\lambda)$ 
describes distribution of probabilities for Poisson distribution
with the parameter $\lambda$ and a random variable $x$

Let us rewrite the density of  Gamma distribution using unconventional
notation

\begin{equation}
f(x,a,\lambda) = 
\displaystyle \frac{a^{x+1}}{\Gamma(x+1)} e^{-a\lambda} \lambda^{x}, 
\end{equation}

\noindent
where $a$ is a scale parameter, $x > -1$ is a shape 
parameter\footnote{A conventional notation for the shape parameter 
is~\cite{6} $r = x + 1,~r > 0$.} and 
$\lambda > 0$ is a random variable. Here the quantities of $x$ 
and $\lambda$ take values from the set of real numbers.
Let us set $a = 1$ and denote $x! = \Gamma(x+1)$, then
a continuous function

\begin{equation}
f(x,\lambda) = \displaystyle \frac{\lambda^x}{x!} e^{-\lambda},~
\lambda > 0,~x > -1 
\end{equation}

\noindent
is the density of  
Gamma distribution with the scale parameter $a = 1$. 

To anticipate a little, it is indicative of the Gamma distribution of
parameter $\lambda$ for the Poisson distribution in case of observed
value $x = \hat x$.

Figure 1 shows the surface described by the function
$f(x,\lambda)$. Smooth behaviour of this function along
$x$ and $\lambda$ (see Fig.2) allows to assume that there is such a function
$\l(\lambda) > -1$, that 

\begin{equation}
\displaystyle \int_{l(\lambda)}^{\infty}{f(x,\lambda)dx} = 1 
\end{equation}

\noindent
for given value of $\lambda$. It means that in this way we introduce continued 
analogue of Poisson distribution with the probability density
$f(x,\lambda) = \displaystyle \frac{\lambda^x}{x!} e^{-\lambda}$ over the
area of the function definition, i.e. for $x \ge l(\lambda)$ 
and $\lambda > 0$. 
The values of the function $f(x,\lambda)$ for integer $x$ 
coincide with corresponding magnitudes in the probabilities distribution
of discrete Poisson distribution.
Dependences of the values of function $\l(\lambda)$, the means and the
variances for the suggested distribution on $\lambda$ have been calculated
by using programme DGQUAD from the library  
CERNLIB~\cite{7} and the results are presented in Table 1.
This Table shows that series of properties of Poisson distribution
$(E\xi = \lambda, D\xi = \lambda)$ take place only if the value of 
the parameter $\lambda > 3$.  

It is appropriate at this point to say that

\begin{equation}
\displaystyle \int_0^{\infty}{f(x,\lambda)dx} = 
\displaystyle \int_0^{\infty}{\frac{\lambda^xe^{-\lambda}}{\Gamma(x+1)}dx} =
e^{-\lambda}\nu(\lambda). 
\end{equation}

\noindent
The function 

\begin{equation}
\nu(\lambda) = 
\displaystyle \int_0^{\infty}{\frac{\lambda^x}{\Gamma(x+1)}dx} 
\end{equation}

\noindent
is well known
and, according to ref.~\cite{8},

\begin{equation}
\nu(\lambda) = 
\displaystyle \sum_{n=-N}^{\infty}{\frac{\lambda^n}{\Gamma(n+1)}}
+ O(|\lambda|^{-N-0.5}) = e^{\lambda} + O(|\lambda|^{-N}) 
\end{equation}

\noindent
if $\lambda \rightarrow \infty,~~|arg \lambda| \le \frac{\pi}{2}$ 
for any integer $N$. Nevertheless we have to use the function $l(\lambda)$
in our calculations in Section 3. We consider it as a mathematical trick
to illustrate a possibility of construction of confidence intervals 
by numerically in the case of real value $\hat x$.

\begin{table}[t]

\small
    \caption{The function $l(\lambda)$,
    mean and variance versus $\lambda$.
}
    \label{tab:Tab.1}

\footnotesize
    \begin{center}
\begin{tabular}{|r|r|r|r|}
\hline
$\lambda$ & $l(\lambda)$ & mean $(E\xi)$ & variance $(D\xi)$ \\ 
\hline
    0.001 &  -0.297 &     -0.138  &  0.024  \\
    0.002 &  -0.314 &     -0.137  &  0.029  \\
    0.005 &  -0.340 &     -0.130  &  0.040  \\
    0.010 &  -0.363 &     -0.120  &  0.052  \\
    0.020 &  -0.388 &     -0.100  &  0.071   \\
    0.050 &  -0.427 &     -0.051  &  0.113    \\
    0.100 &  -0.461 &      0.018  &  0.170    \\
    0.200 &  -0.498 &      0.142  &  0.272    \\
    0.300 &  -0.522 &      0.256  &  0.369    \\
    0.400 &  -0.539 &      0.365  &  0.464   \\
    0.500 &  -0.553 &      0.472  &  0.559    \\
    0.600 &  -0.564 &      0.577  &  0.653    \\
    0.700 &  -0.574 &      0.681  &  0.748    \\
    0.800 &  -0.582 &      0.785  &  0.844    \\
    0.900 &  -0.590 &      0.887  &  0.939    \\
     1.00 &  -0.597 &      0.989  &  1.035    \\
     1.50 &  -0.622 &      1.495  &  1.521    \\
     2.00 &  -0.639 &      1.998  &  2.012    \\
     2.50 &  -0.650 &      2.499  &  2.506    \\
     3.00 &  -0.656 &      3.000  &  3.003    \\
     3.50 &  -0.656 &      3.500  &  3.501   \\
     4.00 &  -0.647 &      4.000  &  3.999    \\
     4.50 &  -0.628 &      4.500  &  4.498    \\
     5.00 &  -0.593 &      5.000  &  4.997    \\
     5.50 &  -0.539 &      5.500  &  5.497    \\
     6.00 &  -0.466 &      6.000  &  5.996    \\
     6.50 &  -0.373 &      6.500  &  6.495    \\
     7.00 &  -0.262 &      7.000  &  6.995    \\
     7.50 &  -0.135 &      7.500  &  7.494    \\
     8.00 &   0.000 &      8.000  &  7.993    \\
     8.50 &   0.000 &      8.500  &  8.496    \\
     9.00 &   0.000 &      9.000  &  8.997    \\
     9.50 &   0.000 &      9.500  &  9.498    \\
     10.0 &   0.000 &      10.00  &  9.999    \\
\hline
\end{tabular}
    \end{center}
\end{table}

In principle, we can numerically transforms the 
function $f(x,\lambda)$ in the interval $x\in(0,1)$ so that 

\noindent
$\displaystyle \int_0^{\infty}{f(x,\lambda)dx} = 1,~~
E\xi = \displaystyle \int_0^{\infty}{xf(x,\lambda)dx} = \lambda,$

\noindent
$D\xi = \displaystyle
\int_0^{\infty}{(x-E\xi)^2f(x,\lambda)dx} = \lambda$

\noindent
for any $\lambda$. In this case we can construct confidence intervals
without introducing of $l(\lambda)$.

In Section 3 assumption about continuous of the function
$f(x,\lambda)$ along the variable $x$ are used for construction
of confidence intervals of a parameter $\lambda$ for any observed $\hat x$.

Let us construct a central confidence intervals for the continued
analogue of Poisson distribution using function $l(\lambda)$.

\section{The Construction of the Confidence Intervals for
Continued Analogue of Poisson Distribution.}

Assume that in the experiment with the fixed integral luminosity
(i.e. a process under study may be considered as a homogeneous process
during given time) the $\hat x$ events of some Poisson process
were observed. It means that we have an experimental estimation
$\hat \lambda(\hat x)$ of the parameter $\lambda$ of Poisson distribution.
We have to construct a confidence interval
$(\hat \lambda_1(\hat x), \hat \lambda_2(\hat x))$, covering
the true value of the parameter $\lambda$ of the distribution under
study with confidence level $1 - \alpha$, where $\alpha$ is a
significance level. It is known from the theory of statistics~\cite{9},
that the mean value of a sample of data is an unbiassed estimation
of mean of distribution under study. In our case the sample consists
of one observation $\hat x$. For the discrete Poisson distribution
the mean coincides with the estimation of parameter value, 
i.e. $\hat \lambda = \hat x$.
This is not true for a small value of $\lambda$ in the considered case
(see Table 1). That is why in order to find the estimation of
$\hat \lambda(\hat x)$ for small value $\hat x$ there is necessary
to introduce correction in accordance with Table 1. Let us construct
the central confidence intervals using conventional method assuming that

\begin{equation}
\displaystyle \int_{\hat x}^{\infty}{f(x,\hat \lambda_1)dx} = \frac{\alpha}{2} 
\end{equation}

\noindent
for the lower bound $\hat \lambda_1$ and

\begin{equation}
\displaystyle
\int_{l(\hat \lambda_2)}^{\hat x}{f(x,\hat \lambda_2)dx} = \frac{\alpha}{2}
\end{equation}

\noindent
for the upper bound $\hat \lambda_2$ of confidence interval.

Figure 3 shows the introduced in the Section 2 distributions with
parameters defined by the bounds of confidence interval 
$(\hat \lambda_1 = 1.638, \hat \lambda_2 = 8.493)$
for the case $\hat x = \hat \lambda = 4$ and the
Gamma distribution with parameters $a = 1$, $x = \hat x = 4$. 
The association between the confidence interval and
the Gamma distribution is seen from this Figure.
The bounds of confidence interval with 90\% confidence level
for the parameter of continued analogue of Poisson distribution for different
observed values $\hat x$ (first column) were calculated and are given 
in the second column of the Table 2.
It is necessary to notice that the confidence level of the constructed
confidence intervals always coincides exactly with the required
confidence level. As it results from Table 2 that the suggested approach 
allows to construct confidence intervals for any real and integer 
values of the observed number of events in the case of the values of
parameter $\lambda > 3$. The Table 2 shows that the left bound of central 
confidence intervals is not equal to zero for small $\hat x$, what is 
not suitable.

Also note that 90\% of the area of Gamma distributions with parameter
$x = \hat x$ are contained inside the constructed 90\% confidence intervals
for observed value $\hat x$. However, for small values of $\hat x$ we have got 
values of the area close to 88\%, i.e. less than 90\%.   

\begin{table}[t]
\small
    \caption{90\% C.L. intervals for the Poisson signal mean $\lambda$
    for total events observed $\hat x$.}
    \label{tab:Tab.2}
\footnotesize
    \begin{center}
\begin{tabular}{|r|rr|rr|rr|}
\hline
    & bounds   & (Section 3) & bounds & (Section 4) & bounds & (ref[1])   \\ 
$\hat x$ &$\hat \lambda_1$&$\hat \lambda_2$&$\hat \lambda_1$&$\hat \lambda_2$ &
$\hat \lambda_1$ & $\hat \lambda_2$ \\
\hline
 0.000 & 0.121E-08 &  2.052 &0.0       &  2.303 & 0.00 & 2.44  \\
 0.001 & 0.205E-08 &  2.054 &0.0       &  2.304 &      &     \\
 0.002 & 0.292E-08 &  2.056 &0.0       &  2.306 &      &     \\
 0.005 & 0.666E-08 &  2.061 &0.0       &  2.311 &      &     \\
 0.01 & 0.307E-07 &  2.076 & 0.0       &  2.320 &      &     \\
 0.02 & 0.218E-06 &  2.098 & 0.0       &  2.337 &      &     \\
 0.05 & 0.765E-05 &  2.166 &  1.66E-05 &  2.389 &      &     \\
 0.10 & 0.137E-03 &  2.275 &  2.23E-05 &  2.474 &      &    \\
 0.20 & 0.186E-02 &  2.490 &  6.65E-05 &  2.642 &      &    \\
 0.30 & 0.696E-02 &  2.692 &  1.49E-04 &  2.806 &      &    \\
 0.40 & 0.161E-01 &  2.891 &  2.60E-03 &  2.969 &      &    \\
 0.50 & 0.295E-01 &  3.084 &  5.44E-03 &  3.129 &      &    \\
 0.60 & 0.466E-01 &  3.269 &  1.35E-02 &  3.290 &      &    \\
 0.70 & 0.673E-01 &  3.450 &  2.63E-02 &  3.452 &      &    \\
 0.80 & 0.911E-01 &  3.629 &  4.04E-02 &  3.611 &      &    \\   
 0.90 & 0.1179     &  3.804 & 6.12E-02 &  3.773 &      &    \\   
  1.0 & 0.1473     & 3.977  & 8.49E-02 &  3.933 & 0.11 & 4.36  \\
  1.5 & 0.3257     & 4.800  & 0.2391   &  4.718 &      &       \\
  2.0 & 0.5429     & 5.582  & 0.4410   &  5.479 & 0.53 & 5.91  \\
  2.5 & 0.7896     & 6.340  & 0.6760   &  6.220 &      &       \\
  3.0 & 1.056      & 7.076  & 0.9284   &  6.937 & 1.10 & 7.42  \\
  3.5 & 1.340      & 7.792  & 1.219    &  7.660 &      &       \\
  4.0 & 1.638      & 8.493  & 1.511    &  8.358 & 1.47 & 8.60  \\
  4.5 & 1.946      & 9.188  & 1.820    &  9.050 &      &       \\
  5.0 & 2.264      & 9.869  & 2.120    &  9.714 & 1.84 & 9.99  \\
  5.5 & 2.590      & 10.55  & 2.453    &  10.39 &      &       \\
  6.0 & 2.924      & 11.21  & 2.775    &  11.05 & 2.21 & 11.47 \\   
  6.5 & 3.264      & 11.87  & 3.126    &  11.72 &      &       \\
  7.0 & 3.609      & 12.53  & 3.473    &  12.38 & 3.56 & 12.53 \\
  7.5 & 3.961      & 13.18  & 3.808    &  13.01 &      &       \\
  8.0 & 4.316      & 13.82  & 4.160    &  13.65 & 3.96 & 13.99 \\
  8.5 & 4.677      & 14.46  & 4.532    &  14.30 &      &       \\
  9.0 & 5.041      & 15.10  & 4.905    &  14.95 & 4.36 & 15.30 \\
  9.5 & 5.406      & 15.73  & 5.252    &  15.56 &      &       \\   
  10. & 5.779      & 16.36  & 5.640    &  16.21 & 5.50 & 16.50 \\   
  20. & 13.65      & 28.49  & 13.50    &  28.33 & 13.55 & 28.52 \\
\hline
\end{tabular}
    \end{center}
\end{table}

\section{Shortest Confidence Intervals for
Parameter of Poisson Distribution.}

As is follow from the formula $(3)$ for $f(x,\lambda)$ (see Fig.3) 
the true value of $\lambda$ of Poisson distribution for the
observed value $\hat x$ has Gamma distribution (at least both
for integer $x \ge 0$ if we consider pure Poisson distribution and
for real $x \ge 0$ if we consider analogue of Poisson distribution)
with the parameters $a = 1$ and $x = \hat x$, i.e. 

\begin{equation}
P(\lambda|\hat x) = 
\displaystyle \frac{\lambda^{\hat x}}{\hat x!} e^{-\lambda}. 
\end{equation}

\noindent
This supposition allows 
to choose confidence interval of a minimum length from all possible
confidence intervals of given confidence level 
without violation of the coverage principle. The bounds of 
minimum length area, 
containing 90\% of the corresponding
Gamma distribution square, were found numerically both for 
integer value of $\hat x$ and for real value of $\hat x$.
We took into account that $x = \hat x$ and found the shortest
90\% confidence interval for the parameter of Poisson distribution.
The results are presented in third column of the Table 2.
For comparison with the results of conventional procedure~\cite{2}
of finding confidence intervals, the results of calculations of
confidence intervals for integer value of $\hat x$~\cite{1} are 
adduced in the Table 2. By this means confidence intervals,
got using Gamma distribution, may be used for real values of $\hat x$.

\section{Conclusion}

In the article the attempt of introducing of continued
analogue of Poisson distribution for the construction of classical
confidence intervals for the parameter $\lambda$ of Poisson distribution
is described. 
Also, in the article is shown that the true value of a parameter of Poisson 
distribution for the observed value $\hat x$ has Gamma distribution 
with the scale parameter $a = 1$ and the shape parameter $x = \hat x$.
Two approaches (with using of function $l(\lambda)$ 
and with using of Gamma distribution) are considered. 
Confidence intervals for different integer and real 
values of number of observed events for Poisson process in the experiment
with given integral luminosity are constructed. The second
approach allows to construct any confidence intervals for a parameter
of Poisson distribution.

\begin{flushleft}
{\large \bf Acknowledgments}
\end{flushleft}

We are grateful to V.A.~Matveev, V.F.~Obraztsov and Fred James
for the interest to this work and for valuable comments. 
We are thankful to S.S.~Bityukov and V.A.~Litvine for useful
discussions. We would like
to thank E.A.Medvedeva for the help in preparing the article.
This work has been supported by RFFI grant 99-02-16956 
and grant INTAS-CERN 377.

\begin{figure}[H]

\centerline{\epsfig{file=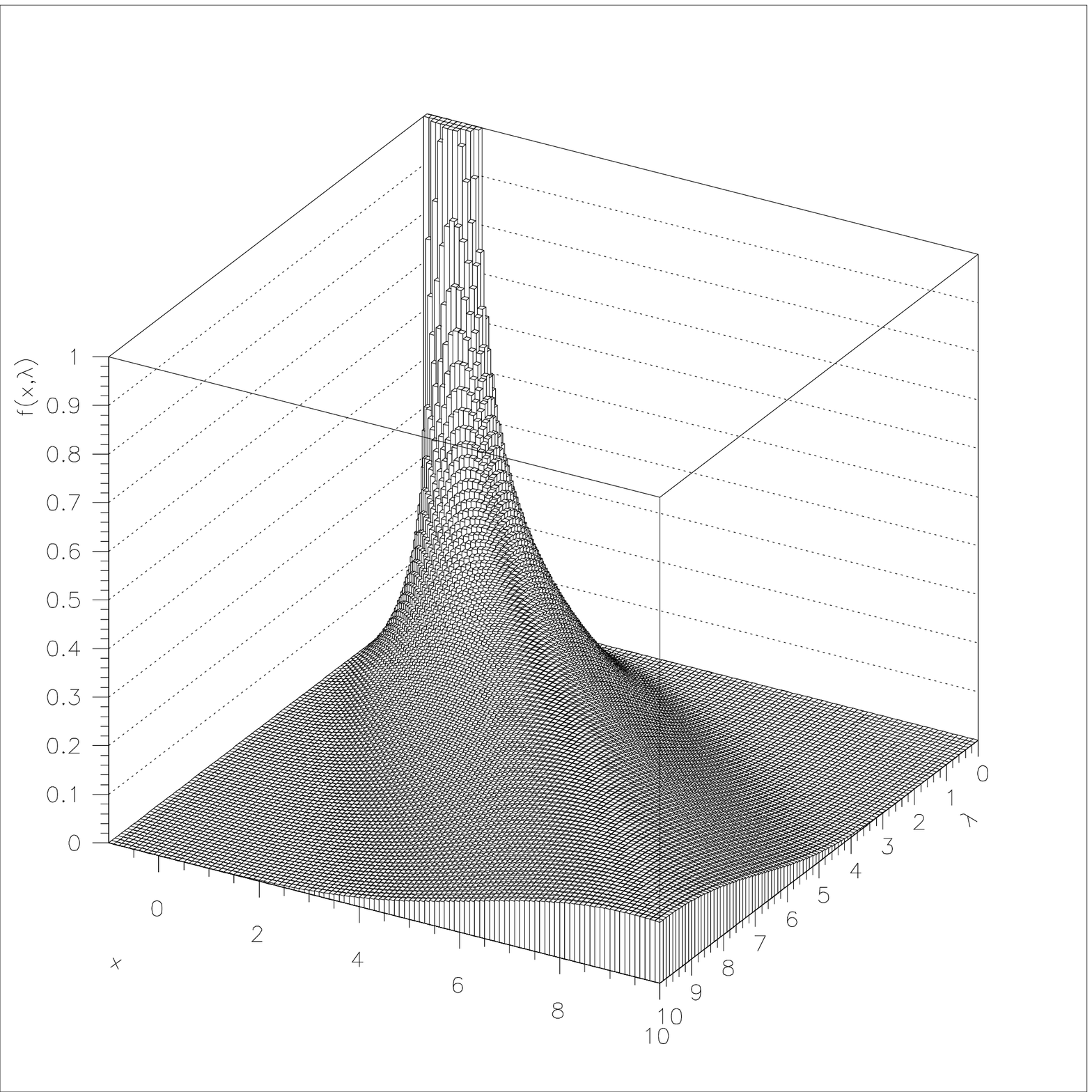,width=14cm}}

\vspace*{-0.4cm}

\caption{\small The behaviour of the function $f(x,\lambda)$ versus
$\lambda$ and $x$ if $f(x,\lambda)<1$.
}
\label{fig.1}

\end{figure}

\begin{figure}[H]

\centerline{\epsfig{file=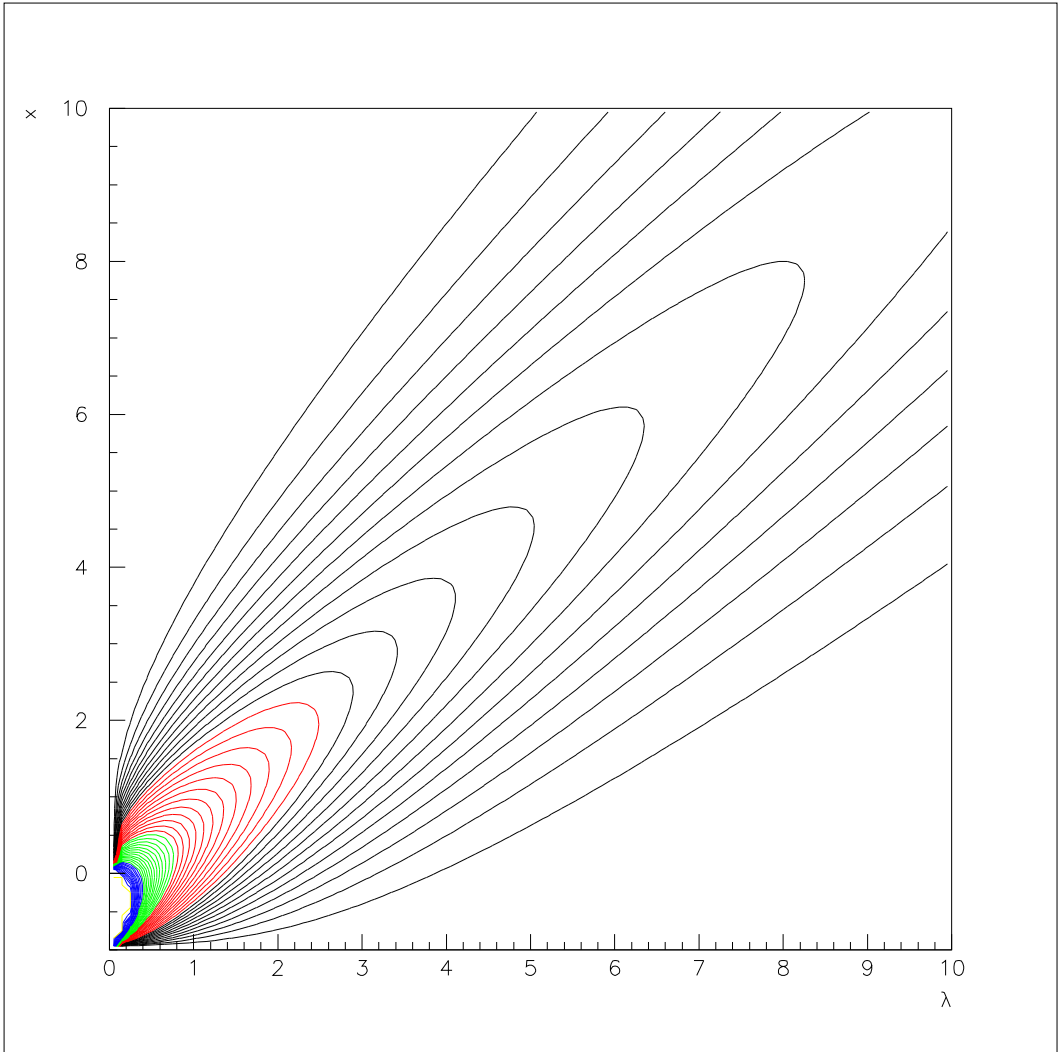,width=14cm}}

\vspace*{-0.4cm}

\caption{\small Two-dimensional representation 
of the function $f(x,\lambda)$ versus
$\lambda$ and $x$ for values $f(x,\lambda) < 1$. 
}
\label{fig.2}

\end{figure}

\begin{figure}[H]

\centerline{\epsfig{file=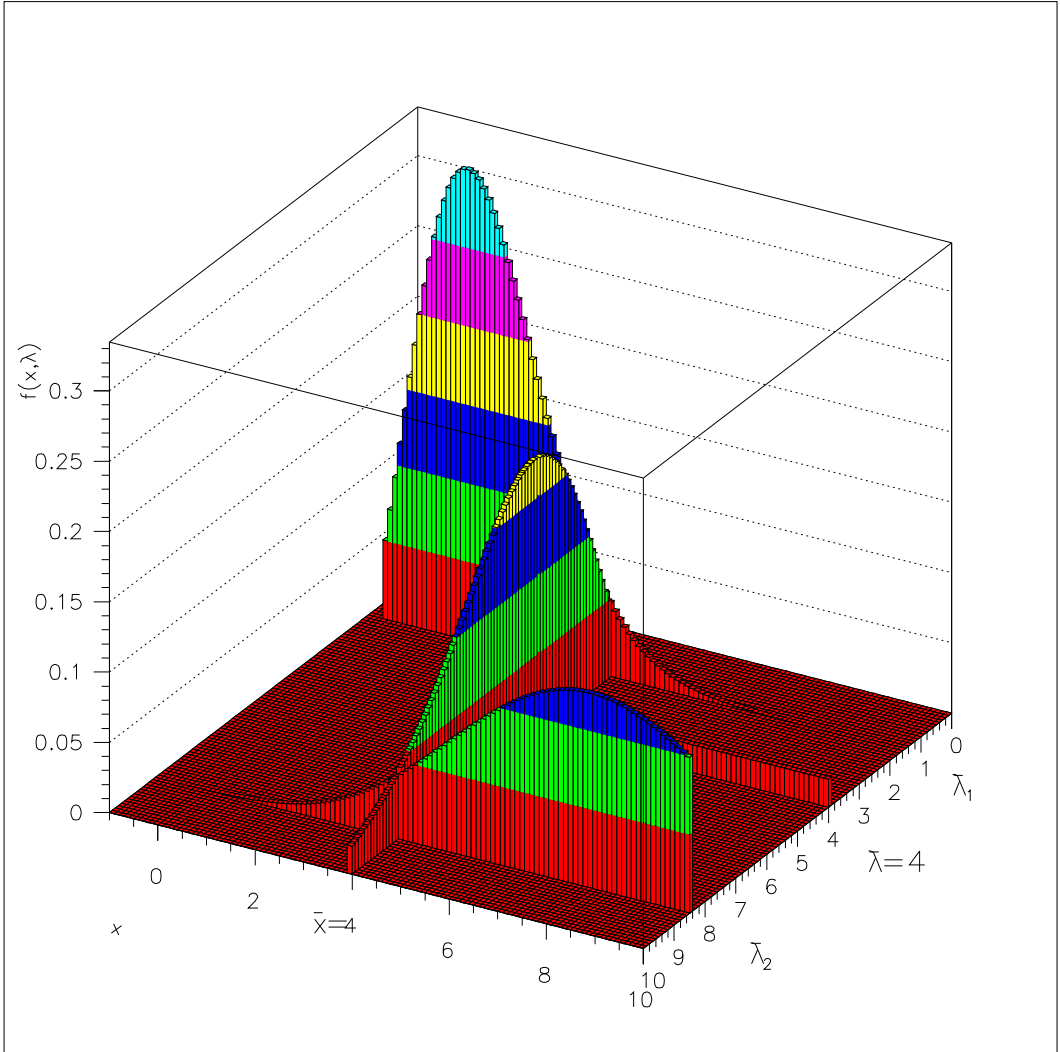,width=14cm}}

\vspace*{-0.4cm}

\caption{\small The probability densities $f(x,\lambda)$ 
of continuous analogue Poisson distribution for $\lambda$'s
determined by the
confidence limits $\hat \lambda_1$ and $\hat \lambda_2$
in case of observed number of events $\hat x = 4$ and the
probability density of Gamma distribution with parameters $a=1$ and
$x=\hat x=4$.}
\label{fig.3}

\end{figure}

\end{document}